\documentclass[aps,prd,twocolumn,amsmath,showpacs,showkeys,floatfix]{revtex4}
\usepackage[dvips]{graphicx}

\newcommand{\CP}{\textit{CP}}
\newcommand{\Ks}{K_{\mbox{\scriptsize S}}}
\newcommand{\KKbar}{$K$-$\bar K$}
\newcommand{\BBbar}{$B$-$\bar B$}
\newcommand{\BBbarq}{$B_{q}$-$\bar B_{q}$}
\newcommand{\BBbard}{$B_{d}$-$\bar B_{d}$}
\newcommand{\BBbars}{$B_{s}$-$\bar B_{s}$}
\newcommand{\Vi}{V_{\,\mbox{\scriptsize\textrm{I}}}}
\newcommand{\Vii}{V_{\,\mbox{\scriptsize\textrm{II}}}}
\newcommand{\Hd}[1]{H_{\Delta#1=2}}
\newcommand{\AJK}{\ensuremath{{\cal A}_{\psi K}}}
\newcommand{\vstrut}[2]{\vrule width 0pt height #1ex depth #2ex}
\renewcommand{\Im}{\mathop{\textrm{Im}}}

\begin{document}
\preprint{UCI-TR-2003-15}
\title{ Effects of Right-handed Gauge Bosons
  in {\boldmath $B$-$\bar{B}$} Mixing and {\boldmath $CP$} Violation}
\author{Dennis Silverman}
\affiliation{Department of Physics and Astronomy, University of California,
  Irvine, CA 92697-4575} 
\author{W. K. Sze} 
\author{Herng Yao} 
\affiliation{Department of Physics, National Taiwan Normal University,
  Taipei, Taiwan 117} 
\date{\today}                                                

\begin{abstract}
In the Left-Right Symmetric Model (LRSM),
box diagrams involving the charged right-handed gauge boson~$W_R$
may affect \BBbar\ mixing as well as $CP$ asymmetries in neutral $B$ decays.
The smallness of the $\epsilon_K$ parameter in the neutral $K$-meson system
places severe constraints on the right-handed quark mixing matrix~$V^R$,
and reduces the number of its effective phases to one.
$W_R$ exchange gives a large contribution to 
\BBbar\ mixing when the mass of the $W_{R}$ boson is up to or higher 
than $8$ TeV, depending on the $V^R$ case, the $B_{d,s}$ meson, and
the asymmetry.  
The allowed regions of the  $CP$ violating asymmetries $\sin{2\beta}$, 
$\sin{\gamma}$,
$\sin{2\alpha}$, and $\sin{2\phi_s}$, as well as $x_s$, are 
calculated as a function of the $W_R$ mass.
The results of the LRSM other than for the well measured $\sin{2\beta}$
show allowable regions that are much broader than that for the Standard Model,
showing that new experiments can indicate a presence of new physics,
or significantly push up the limits on the $W_R$ mass.

\end{abstract} 
                                           

\maketitle                                                   

\section{Introduction}

The Standard Model~(SM) for strong and electroweak interactions has achieved
great success in explaining interactions among the elementary particles.
Nonetheless,
it is generally speculated that we may encounter richer symmetry structures,
such as supersymmetry or larger gauge groups, as we go to higher energy scales.
For example, in the Left-Right Symmetric Model~(LRSM) \cite{mohapatra},
the right-handed quarks have gauge interactions among
themselves, just like their left-handed counterparts,
and such interactions have not yet been seen only because they are 
spontaneously broken at an energy higher than the electroweak scale.
By examining such extended models to the SM,
we may get a better understanding of the mixing and mass scales
whose values are not restricted by any principle we know so far.

$CP$ violation is an excellent realm to look for experimental effects of
the LRSM, because the right handed mixings in LRSM bring in another
$3\times3$ quark mixing matrix $V^R$ which has new phases that can affect
$CP$ violation.  We will see that constraints reduce the number of 
effective new phases to one in the $B_d$ or $B_s$ systems.

The main experimental data giving or constraining $CP$ violation are listed
in Table~\ref{V:ex} \cite{hagiwara}.
Among the data listed in Table~\ref{V:ex}, the $\epsilon_K$ parameter
in the neutral $K$ meson system has been known for a long time,
and until recently was the only direct evidence for \CP~violation.
With the advent of $B$ factories, it is now possible
to investigate $CP$ violating effects in the neutral $B$ meson system as well.
Carter and Sanda \cite{carter} proposed that $CP$ violating effects in
{\BBbarq} ($q=d,s$) mixing can be probed by investigating decays 
of~$B_q$($\bar B_q$) to a \CP~eigenstate~$f$.
The time-dependent \CP~asymmetries would show an oscillatory behavior,
with a characteristic amplitude~${\cal A}_f$.
For example, in the decay $B\to J/\psi\,\Ks$,
the asymmetry $\AJK=\sin{2\beta}$ in the SM.
\begin{table}
\renewcommand{\arraystretch}{1.3}
\caption{\label{V:ex}}%
Experiments and values constraining $V$ 
\begin{ruledtabular}
\begin{tabular}{@{\qquad}cr@{\extracolsep{0pt}}l@{\qquad}}
  $|V_{ub}|$ & $(3.6\pm0.7)$ & ${}\times10^{-3}$ \\
  $|V_{cb}|$ & $(41.2\pm2.0)$ & ${}\times10^{-3}$ \\
  $|\epsilon_K|$ & $(2.282\pm0.017)$ & ${}\times10^{-3}$ \\
  $\Delta M_{B_d}$ & \multicolumn{2}{c}{$0.502\pm0.006\,\textrm{ps}^{-1}$}\\
  $x_s$          & $ > 19.0 \textrm{~at~95\%~CL}$ & \\
  $\sin{2\beta}$ & $   0.735\pm 0.056$            &  
\end{tabular}
\end{ruledtabular}
\end{table}

In this paper we investigate the possibility of larger ranges for 
these $CP$ violating asymmetries in processes involving $W_R$ bosons
and $V^R$ mixing in the LRSM.  We use two models for the right handed
coupling matrix which allow large couplings for box diagrams with
two $t$ quark sides containing one $W_R$ and one $W_L$ exchange, 
allowing effects in $B_s$ mesons (case I),
or $B_d$ mesons (case II).  For $B_s$ mesons, 
the $t$ quark right hand couples to $s$
as well as to $b$ quarks (case I), but does not also couple to $d$
quarks, in order to minimize the right handed effects in $K$ meson 
mixing or $\epsilon_K$.  For $B_d$ mesons, 
the $t$ quark right hand couples to
$d$ as well as to $b$ quarks (case I), but not also to $s$ quarks, 
in order to minimize the effect in $\epsilon_K$.  
Although several phases and one
mixing angle are present in each case, there is only one phase
combination which contributes to $CP$ violating asymmetries, and
we vary it over all values.   

We find significant effects from box diagrams containing both 
$W_L$ and $W_R$ exchanges for the $W_R$ mass up to $8$ TeV or higher,
for both coupling cases I and II.
Experiments on the $CP$ violating asymmetries and on $x_s$
may find results outside the SM range, which would indicate new
physics, such as the LRSM.

In the next section we will give a brief review of $CP$
violation in the SM, and then in Sec. III of mixing in the LRSM,
and describe its implication for neutral $B$ meson physics.
Apparently there are a few phases in the right-handed quark mixing matrix~$V^R$
which can contribute to \CP~violation in the $B$ sector, but the constraints
imposed by~$\epsilon_K$ reduce the number of effective phases to only one,
as discussed in Sec. IV.
The effects of $W_R$ on \BBbar\ mixing is given in Sec. V, and its
magnitudes shown in Sec. VI.  Effects of $W_R$ on specific 
$CP$ violating asymmetries of
$\sin{2\beta}$, $\sin{\gamma}$, $\sin{2\alpha}$, and 
$\sin{2\phi_s}$ 
are presented in Sections VII, VIII, IX, and X, respectively.
$x_s$ is covered in Section XI, and the conclusions are 
summarized in Section XII.

\section{{\boldmath $CP$} Violation in the Standard Model}

In the SM with three generations of quarks,
all \CP-violating effects come from the $3\times3$ left-handed quark mixing
matrix~$V\equiv V^L$ (the CKM matrix) \cite{CKM}.
There are many possible ways to parameterize the phases appearing in~$V$,
but there is only one single independent re-parameterization invariant measure.
For example, in Wolfenstein's parameterization \cite{wolfenstein}:
\begin{equation}
\label{ckm-matrix:wolf}
\renewcommand{\arraystretch}{1.3}
  V = \begin{pmatrix}
    1-\frac{1}{2}\lambda^2   & \lambda &  A\lambda^{3}(\rho - i\eta)  \\ 
    -\lambda                   & 1 - \frac{1}{2}\lambda^{2} & A\lambda^2\\
    A\lambda^{3}(1-\rho-i\eta) &  -A\lambda^{2}             & 1
  \end{pmatrix}.
\end{equation}
the \CP~violating phases are assigned to the smallest elements
$V_{ub} \equiv |V_{ub}|e^{-i\gamma}$ and $V_{td} \equiv |V_{td}|e^{-i\beta}$,
and therefore, the phase angles $\beta$ and~$\gamma$ can be sizable
in spite of the tiny intrinsic \CP~violation.
The \CP~violating parameter can then be taken as the area of the ($d$-$b$)
unitarity triangle, which is half of Jarlskog's parameter~$J$ \cite{jarlskog}.

The parameter $\lambda\equiv|V_{us}|=0.221$ in Eq.~(\ref{ckm-matrix:wolf})
is quite accurately known,
and we will disregard its small uncertainty in subsequent analysis.
The other three parameters are presently known,
with larger uncertainties, to be of order~1.

On the other hand, if we normalize the base of the unitarity triangle
to unit length, as is usually done,
the other two sides of the triangle would be given by
$|V_{ub}|/\lambda\,|V_{cb}|$ and $|V_{td}|/\lambda\,|V_{cb}|$,
with~$\beta$ and $\gamma$ as their respective opposite angles.
Thus the first four data listed in Table~\ref{V:ex} were sufficient to 
give estimates for
the angles $\beta$ and~$\gamma$.
Values of $\sin{2\beta}$ obtained by earlier theoretical analyses were
\begin{equation}
\label{sin2beta:th}
\sin{2\beta} = \begin{cases} 0.75\pm 0.06 & \hbox{\cite{cara}},\\
                   0.73 \pm 0.20 & \hbox{\cite{ttt}}.
             \end{cases}
\end{equation}
The experimental value of~\AJK\ can then be compared with these theoretical
predictions.

Measurements of the \CP~violating asymmetry in the decay $B\to J/\psi\,\Ks$
now exist to high precision.
The values obtained by BaBar \cite{babar}, Belle \cite{belle},
and CDF \cite{cdf} are
\begin{equation}
\AJK = \begin{cases} 0.741 \pm 0.067 \pm 0.034 & (\rm BaBar), \\
                      0.719 \pm 0.074 \pm 0.035 & (\rm Belle), \\
                      0.79 \pm 0.42 & (\rm CDF).
\label{AJK:ex0}
       \end{cases}
\end{equation}
These give the world-averaged value
\begin{equation}
\label{AJK:ex}
\AJK = 0.735 \pm 0.056.
\end{equation}
We see that there is excellent agreement between Eq.~(\ref{sin2beta:th})
and Eq.~(\ref{AJK:ex}).
Besides~$\beta$, the SM also has definite predictions for other $CP$
violating asymmetries.

\section{The Left-Right Symmetric Model}

In the LRSM, one assumes that the Lagrangian for the elementary particles obeys
(apart from the $SU(3)_c$ for strong interaction)
an $SU(2)_{L}\times SU(2)_{R}\times U(1)_{B-L}$ symmetry which is
spontaneously broken at a scale~$v'$ to the electroweak gauge group
$SU(2)_{L}\times U(1)_{Y}$.
There would appear extra charged gauge bosons~$W_R^\pm$ which mediate coupling
with strength~$g_R$ to the right-handed quarks.
The mixing matrix~$V^R$ among the latter is in general different from~$V^L$.
For models which are relevant to hadron-scale physics, $v'$ would not be
much higher than the electroweak scale $v=250\,\textrm{GeV}$, so that
the $W_R$ mass $M_R=\frac{1}{2}g_R v'$ would not be too large either.
Beall, Bander, and Soni \cite{beall} showed that with the 
manifest LR symmetry mixing matrix
of $V^R = V^L$ that the mass of the $W_R$ must exceed
$1.6$ TeV to not overly affect $\Delta M_K$.  We have calculated that
satisfying the constraints of Table I with manifest LR symmetry
gives a 2-$\sigma$ lower limit on $M_R$ of $1.5$ TeV.  
Olness and Ebel \cite{olness} pointed out that $W_R$
can have sub-TeV mass if $V^R$ takes on some specific forms.
The detailed analysis of Langacker and Sankar \cite{langacker} led to
a similar conclusion, namely, that $M_R$ can attain a lower limit
of about 300~GeV if $V^R$ assumes one of the following two forms:
\begin{subequations}
\setlength{\jot}{1.5ex}
\label{Vi:Vii} 
\begin{align}
  \Vi^{R} & = e^{i\omega} \begin{pmatrix}
    1 & 0 & 0 \\
    0 & \hphantom-ce^{i\tau}  & s e^{i\sigma} \\
    0 & -se^{i\phi} & ce^{i\chi} \end{pmatrix}, \\
  \Vii^{R} & = e^{i\omega} \begin{pmatrix}
    0 & 1 & 0 \\
   \hphantom-ce^{i\tau}  & 0 & se^{i\sigma} \\
   -se^{i\phi} & 0 & ce^{i\chi} \end{pmatrix}
  = \Vi^{R} \, \begin{pmatrix}
    0 & 1 & 0 \\
    1 & 0 & 0 \\
    0 & 0 & 1 \end{pmatrix},\quad
\end{align}
\end{subequations}
where for brevity we denote $s\equiv\sin\theta_R$ and $c\equiv\cos\theta_R$.
In what follows, the cases where $V^R=\Vi^R$ and $V^R=\Vii^R$
will be denoted as case~I and case~II, respectively.
Unitarity of~$V^R$ implies that
\begin{equation}
  \tau+\chi = \phi+\sigma \label{phases}
\end{equation}
for both cases.

It has been pointed out that both cases for $V^R$ can lead to sizeable
contributions to \CP~violations for $K$ and $B$ mesons \cite{london}.
For the neutral $B$~mesons, deviation from SM~predictions can occur
through modified \BBbar\ mixing effects, where we have additional box diagrams
with one~$W_L$ and one~$W_R$ exchange \cite{nir}.
In what follows, we will examine how these ``indirect'' effects gives the
corrections to SM~predictions for the \KKbar\ and \BBbar\ systems.
While direct \CP~violating effects involve long distance effects and
final state interactions which are not very well understood, the
long distance effects in box-diagrams have been extensively studied by various
groups with methods such as the $1/N$~expansion or lattice gauge calculations,
and are conveniently summarized in terms of the so-called bag-parameters.

It is clear from Eqs.~(\ref{Vi:Vii}) 
that interchanging the roles played by $d_R$ and~$s_R$
means swapping $\Vi$ and~$\Vii$.
Specifically, since $\Vi$ has non-zero mixing angle factors in the second
and third ($s$ and~$b$) columns only, we expect that case~I will give
substantial corrections to the SM~prediction for the {\BBbars} system,
but not for the \BBbard\ one.
The reverse is true for case~II.
These matrices can also be derived if one first looks for those with
large right handed coupling of
$t$ or $c$ quarks in box diagrams of mixings for $B_d$ or $B_s$,
and then constrains them to not have right handed $t$ or $c$ couplings in
$\epsilon_K$.  The $u$ quark right handed couplings do not count 
in the $B$ systems, due to the
small $u$ quark mass, and are conveniently left out of the mixing
matrix except where needed for unitarity.


\section{Constraints from the $\mathbf{\epsilon_{K}}$ Parameter}

The parameter $\epsilon_{K}$ for \CP\ violations in the \KKbar\ system
is given as $\epsilon_{K} \approx {\rm Im} {\langle  K^{0}| H_{\Delta S = 2} |
 {\bar K}^{0} \rangle} / {\sqrt 2} \Delta m_{K}$.
In the SM, the operator $H_{\Delta S =2}$ is given by the box diagrams
with $W_L$-$W_L$ exchanges \cite{buras}:
\begin{widetext}
\begin{equation}
  \Hd{S} = \frac{G_{F}^{2} M_L^2}{4\pi^{2}}
    [ \eta_{cc}S(x_{c})\zeta_c^2 + \eta_{tt}S(x_{t})
    \zeta_t^2 + 2\eta_{ct}S(x_{c}, x_{t})\zeta_c\zeta_t ]
    (\bar d_Ls_L)(\bar d_Ls_L) + \textrm{h.~c.}  \label{H-ll}
\end{equation}
Here $M_L$ is the $W_L$~mass, $x_i = m_i^2/M_L^2$,
and $\zeta_i \equiv \zeta_i^{LL} = {V_{id}^L}^*V_{is}^L$.
The QCD correction factors are $\eta_{cc} = 1.38, \eta_{tt} = 0.59$, and 
$\eta_{ct} = 0.47$ \cite{herrlich},
and the phase space factors \cite{buras} are
\begin{subequations}
\label{loop-fcn:ll}
\begin{eqnarray}
  S(x) & = & x \left[\frac{1}{4} + \frac{9}{4(1-x) } - \frac{3}{2(1-x)^{2} }
    \right] -\frac{3}{2} (\frac{x}{1-x })^{3} \ln{x}, \\
  S(x_{c},x_{t}) & = & x_{c} \left[\ln\frac{x_{t}}{x_{c}} - 3*\frac{x_{t}}
    {4(1-x_{t})} (1+\frac{x_{t}}{1-x_{t} }\ln{x_{t}})\right].
\end{eqnarray}
\end{subequations}
The effective Hamiltonian in Eq.~(\ref{H-ll}) gives
\begin{equation}
  \epsilon_{K} = \frac{G_{F}^{2}M_{L}^{2} (f_{K}^{2} B_{K}) m_{K}} 
    {12\sqrt{2} \pi^{2} \Delta m_{K} } 
    [ \eta_{cc}S(x_{c})I_{cc} + \eta_{tt}S(x_{t})
    I_{tt} + 2\eta_{ct}S(x_{c}, x_{t})I_{ct} ],
\end{equation}
\end{widetext}
where $I_{ij} \equiv \Im(\zeta_i\zeta_j)$.

In the LRSM,
there are additional contributions to the operator $\Hd{S}$
in which one or both of the~$W_L$ in the box diagram are replaced by the~$W_R$.
For the mixing matrices as given by Eq.~(\ref{Vi:Vii}),
there are no contributions from $W_R$-$W_R$ exchanges,
since the factors~$\zeta_i^{RR} = {V_{id}^R}^*V_{is}^R$ all vanish.
On the other hand,
the $W_L$-$W_R$ exchanges give an additional piece 
$\delta \Hd{S} \equiv H^{LR}$ to $\Hd{S}$ \cite{ecker}:
\begin{widetext}
\begin{equation}
  H^{LR} = \frac{2G^{2}_{F}M_L^{2}}{\pi^{2} }(\frac{g_R}{g_L})^2 \beta_R
    \sum_{i,j=u,c,t}\zeta_{i}^{LR}\zeta_{j}^{RL} 
    J(x_i,x_j,\beta_R)({\bar d_{R}}s_{L})({\bar d_{L}}s_{R}),
\end{equation}
where $\zeta ^{LR}_{i} =V^{L*}_{id}V^{R}_{is}$
and $\beta_R = (M_L/M_R)^{2}$.
Moreover, the loop functions are given by
\begin{equation}
  J(x_i,x_j,\beta_R) \equiv \frac{\sqrt{x_{i}x_{j}}}{4 } [(4\eta_{ij}^{(1)}
    + \eta_{ij}^{(2)}x_{i}x_{j} \beta_R)J_{1}(x_{i},x_{j},\beta_R)
    - (\eta_{ij}^{(3)}+\eta_{ij}^{(4)} \beta_R)J_{2}(x_{i},x_{j},\beta_R)],
  \label{loop-fcn:lr}
\end{equation}
with
\begin{subequations}
  \label{loop-fcn:lr1}
\begin{eqnarray}
  J_{1}(x_{i},x_{j},\beta_R) &=& \frac{x_{i}\ln{x_{i}}}{(1-x_{i})(1-x_{i}
    \beta_R) (x_{i}-x_{j})} +(i\leftrightarrow j)-\frac{\beta_R \ln{\beta_R}}
    {(1-\beta_R) (1-x_{i} \beta_R)(1-x_{j} \beta_R)},
  \\
  J_{2}(x_{i},x_{j},\beta_R) &=& \frac{x_{i}^{2}\ln{x_{i}}}{(1-x_{i})
    (1-x_{i} \beta_R) (x_{i}-x_{j})} +(i\leftrightarrow j) -
    \frac{\ln{\beta_R}}{(1-\beta_R) (1-x_{i} \beta_R)(1-x_{j} \beta_R)}.
\end{eqnarray}
\end{subequations}
\end{widetext}
The $\eta^{(1)-(4)}_{ij}$ are QCD correction factors,
whose explicit forms are given in Ref.~\cite{ecker}.

The contribution of~$H^{LR}$ to~$\epsilon_K$ is given by terms
containing the factors $\Im(\zeta_i^{LR}\zeta_j^{RL})$,
which involves parameters from both~$V^L$ and~$V^R$.
The choices of $V^R$ in Eqs.\ (\ref{Vi:Vii}) eliminate the
$W_L$-$W_R$ box diagrams with the $(t,t)$, $(c,t)$, $(t,c)$, and
$(c,c)$ quark sides.
For leading terms in~$\lambda$,
the non-vanishing mixing angle factors are given in Table~\ref{H-lr:xi},
with $(u,t)$ and $(u,c)$ sides,
and the $W_L$-$W_R$ diagrams are suppressed by $m_u/m_t$ or 
$m_u m_c/m_t^2$, respectively.
\begin{table}
\caption{\label{H-lr:xi}The non-vanishing mixing-angle factors
  $X_{ij} = X_{ji} \equiv \Im(\zeta_i^{LR}\zeta_j^{RL}) + (i\leftrightarrow j)$
  in~$H_{\Delta S=2}^{LR}$.\vstrut{0}{2}}
\begin{ruledtabular}
\begin{tabular}{lcc}
  \vstrut{0}{1.3} & $X_{uc}$ & $X_{ut}$ \\ \hline
  case~I & $-\lambda^2 c \sin\tau$ &
    $-A\lambda^4 \{(1-\rho)^2+\eta^2\}^{1/2} \cdot s \sin(\beta+\phi)$ \\
  case~II & $-c \sin\tau$ & $-A\lambda^2 s \sin\phi$
\end{tabular}
\end{ruledtabular}
\end{table}

The mixing factors shown in Table~\ref{H-lr:xi} have generic values much larger
than their counterparts in the~SM,
which are suppressed by the factors $\lambda^{10}$ for $(t,t)$,
$\lambda^6 m_c/m_t$ for $(c,t)$,
or $\lambda^2 m_c^2/m_t^2$ for $(c,c)$.
Hence we see that constraints need to be imposed on them
for low mass $W_R$ in order that the experimental value
$\left| \epsilon_K \right| = (2.28 \pm 0.02) \times 10^{-3}$ not be exceeded.
These constraints are readily read off from Table~\ref{H-lr:xi} as
\cite{kurimoto}:
\begin{subequations}
\label{constraints}
\begin{eqnarray}
  \textrm{case~I} & : & \quad \tau\approx0,\quad \phi\approx-\beta. \\
  \textrm{case~II} & : & \quad \tau\approx0,\quad \phi\approx0.
\end{eqnarray}
\end{subequations}
The constraint conditions also have solutions which can have $\pi$ 
added separately to the values above.  The important effects of
these extra $\pi$ choices can be included by allowing $\theta_R$ to vary from
$0$ to $\pi$. 
Taking these constraints and the unitarity condition Eq.~(\ref{phases})
into account, Eqs.~(\ref{Vi:Vii}) reduce to
\begin{subequations}
  \label{Vi:Vii:2}
\begin{eqnarray}
  \Vi^{R} &= &e^{i\omega} \begin{pmatrix}
    1 & 0 & 0 \\
    0 & c & s e^{i\sigma} \\
    0 &-se^{-i\beta} & ce^{i(\sigma-\beta)} \end{pmatrix}, \\
  \Vii^{R} &= &e^{i\omega} \begin{pmatrix}
    0 & 1 & 0 \\
    c & 0 & se^{i\sigma} \\
   -s & 0 & ce^{i\sigma} \end{pmatrix}
\end{eqnarray}
\end{subequations}
The case II here is more general than that considered in a previous
paper, Ref.\ \cite{silvermanyao}, 
since it now has a phase and an angle parameter.
The previous case II~\cite{silvermanyao} is equivalent to the
special value $c = 0$ here, with only a phase parameter.
The overall phase~$\omega$ does not appear in the \KKbar\ or \BBbar\ diagrams.
Thus in the cases there is only one relevant free phase left,
which we have taken to be~$\sigma$.
Together with the effective right-handed gauge boson mass
$M_R$ and the mixing angle~$\theta_R$,
we have three more parameters in addition to those in the~SM,
or two more mixing parameters, to the four in the SM.

For large $M_R$ where the constraints Eqs.~(\ref{constraints}) are not 
as stringent, the leading $(t,t)$ LR contributions from Eqs.~(\ref{Vi:Vii})
for $B_d$ and $B_s$ systems
still have only one arbitrary phase, $(\sigma-\tau)$, whose variation 
through all values is included in the
form of Eqs.~(\ref{Vi:Vii:2}), thus maintaining the generality of the 
analysis without the specific form of Eqs.~(\ref{Vi:Vii:2}).

\section{ {\boldmath $B^0$-$\bar{B}^0$ } Mixing in the LRSM} 

The mixing effect in the $B^0_q$-${\bar B^0_q}$ system is given by
\begin{equation}
  x_{q} \equiv \frac{\Delta m}{\Gamma}\bigg|_{B_q} =
    2\tau_{B_{q}}|M_{q,12}|,
\end{equation}
where $q = d$ or $s$,
and $M_{q,12}$ is the dispersive part of the mixing matrix element 
$\langle B_q^{0} | \Hd{B} | \bar{B}_q^{0} \rangle$.
The operator $\Hd{B}$ is similar in form to $\Hd{S}$.
In the~SM,
it is dominated by the box diagram with two internal $t$-quarks
\begin{eqnarray}
  \label{HB-ll}
  \Hd{B} & = & \frac{G_{F}^{2} M_L^2}{4\pi^{2}}
    \eta_{tt} S(x_{t}) \sum_{q=d,s} (V_{tq}^{L*})^2
    (\bar q_Lb_L)(\bar q_Lb_L) \nonumber \\
  & & \qquad + \textrm{h.~c.},
\end{eqnarray}
where $S(x_{t})$ is defined in
Eq. (8a) and the QCD correction factor $\eta_{tt} = 0.59$ in this case.
Eq.~(\ref{HB-ll}) then gives
\begin{equation}
  \label{M_q12^LL}
  M_{q,12}^{LL} = \frac{G^{2}_{F}M_L^{2}}{12\pi^{2} } 
    (f_{B}^{2}B_{B}^{\vphantom{2}}) 
    m_{B}\eta_{tt}S(x_{t}) (V_{tq}^{L*})^{2},
\end{equation}
where we have used the fact that $V_{tb}^L \approx 1$.
The evaluation of the hadronically uncertain factor
$f_{B}^{2}B_{B}^{\vphantom{2}}$ has been the subject of much work,
and recent lattice results are summarized in Ref.\ \cite{becirevic},
giving
\begin{equation}
  f_{B_{q}}^{\vphantom{1/}}B_{B_{q}}^{1/2} =
  \begin{cases}
    228 \pm 32\, \textrm{MeV}, & \hbox{for }q=d, \\
    276 \pm 36\, \textrm{MeV}, & \hbox{for }q=s,
  \end{cases}
\end{equation}
and also $\hat{B}_K = 0.86 \pm 0.13$.

In the LRSM, $\Hd{B}$ receives additional contribution
when one or both of the~$W_L$ in the box diagrams are replaced by~$W_R$,
just as for $\Hd{S}$ in the \KKbar\ system.
As a result we can write
\begin{equation}
  M_{q,12} = M_{q,12}^{LL} + M_{q,12}^{LR} + M_{q,12}^{RR},
\end{equation}
where the element $M_{q,12}^{RR}$ is essentially the same as $M_{q,12}^{LL}$
as given by Eq.~(\ref{M_q12^LL}) but with the replacement $L \to R$ everywhere
(and of course we must retain $V_{tb}^R$, which need not equal to~1):
\begin{eqnarray}
  M_{q,12}^{RR} & = & \frac{G^{2}_{F}M_L^{2}}{12\pi^{2} }
     (\frac{g_R}{g_L})^4 (f_{B}^{2}B_{B}^{\vphantom{2}})m_{B}
    \beta_R \eta_{tt}S(x_{tR})  \cdot{} \nonumber \\
    & & \quad (V_{tq}^{R*}V_{tb}^{R})^{2},  
\end{eqnarray}
where $x_{tR} = (m_t/M_R)^2$.
In case I, this  vanishes in \BBbard\ mixing because $V_{td}^{R} = 0$,
but it has a contribution to \BBbars\ mixing
due to the non-zero values of $V_{ts}^{R}$ and $V_{tb}^{R}$.
In case II, this contributes to \BBbard\ mixing, but vanishes
for \BBbars\ mixing since $V^R_{ts} = 0$ there.

On the other hand, the matrix element $M_{q,12}^{LR}$ is 
\begin{eqnarray}
  \label{Mq12:lr}
  M_{q,12}^{LR} &= &
    \frac{G^{2}_{F}M_L^{2}}{2\pi^{2} } (\frac{g_R}{g_L})^2 
    (f_{B}^{2}B_{B}) (\frac{m_{B}}{m_{b}})^{2}
    m_{B}  \beta_R \cdot{} \nonumber \\
  & & \quad \sum_{i,j=u,c,t}\xi_{q,i}^{LR}\xi_{q,j}^{RL}
    J(x_i,x_j,\beta_R),
\end{eqnarray}
where $\xi^{LR}_{q,i} =V^{L*}_{iq}V^{R}_{ib}$,
$\xi^{RL}_{q,j} =V^{R*}_{jq}V^{L}_{jb}$,
and the function $J(x_i,x_j,\beta_R)$ is defined
in Eqs.~(\ref{loop-fcn:lr})--(\ref{loop-fcn:lr1}).
The QCD-factors $\eta^{(1)-(4)}_{ij}$ in $J(x_i,x_j,\beta_R)$,
like all RG correction factors,
depend on the relevant mass scales only logarithmically,
and hence are relatively insensitive to changes of these masses.
Their values at $M_R=2.5$ TeV and $m_{t} = 175$ GeV,
at the scale $\mu=4.5$ GeV, are given in Table~\ref{qcd-factors:lr} \cite{yao}.
These will be the values we use for subsequent analyses.
We also calculate plots for the left-right symmetry limit $g_R = g_L$.
\begin{table}
\caption{\label{qcd-factors:lr}Values of the QCD-factors $\eta^{(a)}_{ij}$
  in $J(x_i, x_j, \beta_R)$.\vstrut{0}{2}}
\begin{ruledtabular}
\begin{tabular}{@{\quad}rccc@{\quad}}
  & \vstrut{0}{1.3}
    $\eta^{(a)}_{cc}$ & $\eta^{(a)}_{ct}$ & $\eta^{(a)}_{tt}$ \\ \hline
  $a=1$ & 0.61 & 1.27 & 1.98 \\
     2  & 0.04 & 0.27 & 0.75 \\
     3  & 0.55 & 1.03 & 1.93 \\
     4  & 0.45 & 0.84 & 1.58 \\
\end{tabular}
\end{ruledtabular}
\end{table}

In the summation in Eq.~(\ref{Mq12:lr}),
only terms which involve at least one $t$~quark need be considered,
mainly due to the quark mass factors $\sqrt{x_{i}x_{j}}$.
The other terms amount to at most $10^{-3}$ in magnitude
relative to these dominant terms.

To statistically weight the $V^L$~\cite{group} and $V^R$ 
matrix elements angles $s_{23}$, $s_{13}$, and $s$, and phases $\delta$ and 
$\sigma$, we apply six present experimental values, which are those for
$|V_{cb}|$, $|V_{ub}/V_{cb}|=0.087\pm0.017$, 
$\epsilon_{K}$ in the neutral $K$ system, 
\BBbard\ mixing with $\Delta M_{B_{d}}$, (which are listed in Table I), 
as well as the probability of each calculated $x_s$
from the LEPBOSC data average~\cite{lep}, 
and $\sin{2\beta}$ as given in Eq.~(\ref{AJK:ex}).
Complete sets of $V^L$ and $V^R$  angles $s_{23}$, $s_{13}$, $s$, 
and phases $\delta$ and  
and $\sigma$ are generated, and the results of fitting the experiments
are evaluated for each set using $\chi^2$~\cite{silverman}.
$\chi^2$ is formulated as
\begin{equation}
\chi^2 = \sum_{i} \frac{(f_i(s_{23}, s_{13}, \delta, s, \sigma) 
- \langle f_i \rangle )^2}{\sigma_i^2},
\end{equation}
where $\langle f_i\rangle$ and $\sigma_i$ are the  experimental central values 
and deviations for 
$|V_{cb}|$, $|V_{ub}/V_{cb}|$, $\epsilon_{K}$, $\Delta M_{B_{d}}$,
$\sin{2\beta}$, and in $1-A$ for each calculated $x_s$, and 
$f_i(s_{23}, s_{13}, \delta, s, \sigma)$ are 
the corresponding values evaluated in the LRSM cases.  

\section{The Size of the LRSM Contributions to {\boldmath $B$} Mixing}

\subsection { $B_{d}$ systems}

In case I, the loop functions in the $W_L$-$W_R$ box diagrams as given 
by Eq.\ (\ref{Mq12:lr}) are,
like their SM~counterparts, increasing functions of the quark masses.
Due to the vanishingly small mass of the $u$~quark,
only contributions from the $c$- and $t$-exchanges need be considered.
It turns out that, for $\Vi^R$ of the form as given in Eq.~(\ref{Vi:Vii}a),
the mixing angle factors $\xi_{d,c}^{RL}$ and~$\xi_{d,t}^{RL}$
both vanish.
Hence there is essentially no effect of~$W_R$ on $B_d$-$\bar B_d$ mixing
in case I.

\begin{figure}
\noindent\raisebox{30ex}%
  {$\displaystyle\left|\frac{M_{d,12}^{LR}}{M_{d,12}^{LL}}\right|$}
\includegraphics[scale=0.45]{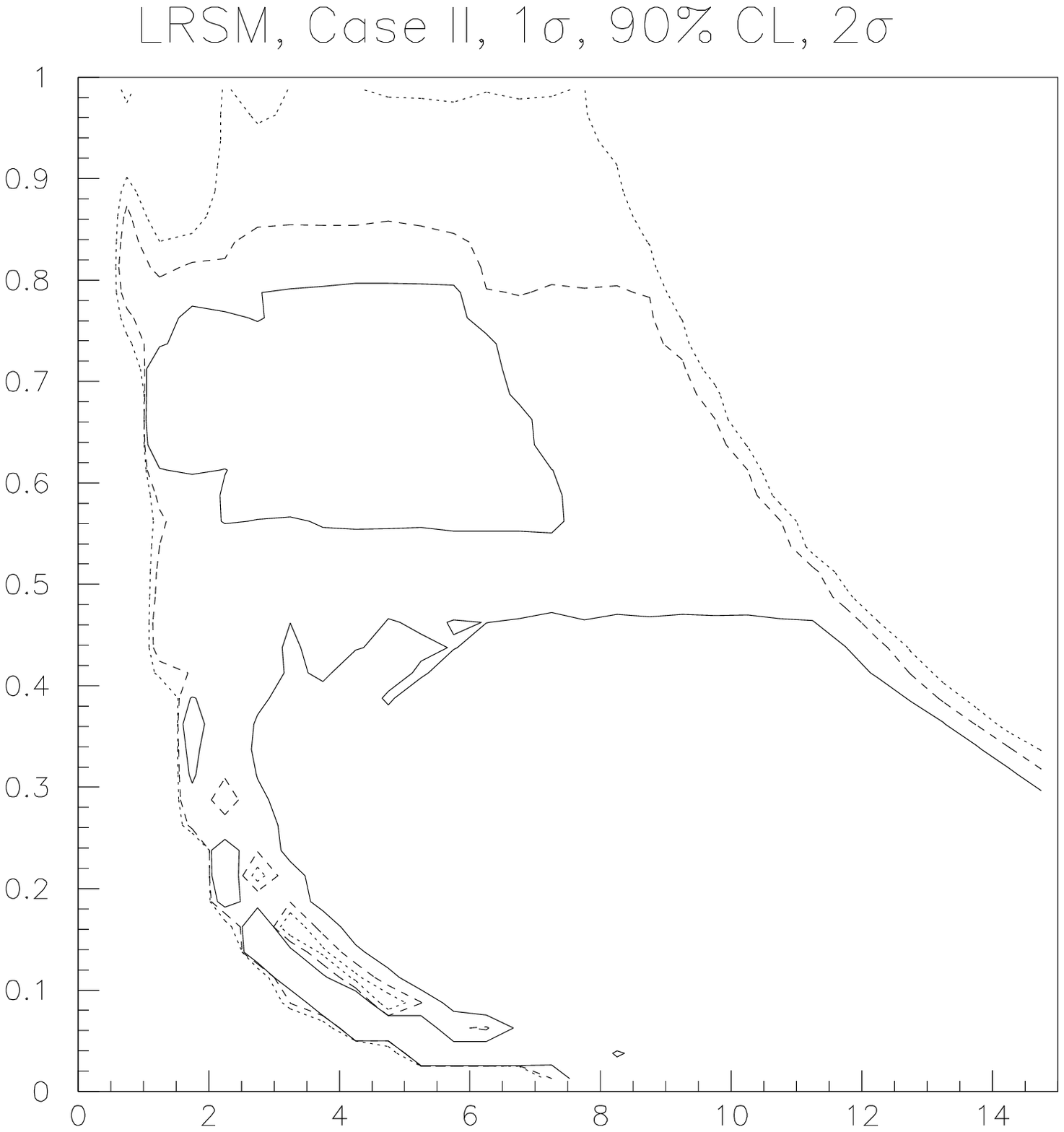}
\hspace*{5em}\mbox{$M_R$ (TeV)}
\caption{\label{relMd12:CaseII}%
  Plot of $\left|M_{d,12}^{LR}/M_{d,12}^{LL}\right|$ as a function
  of $M_R$ for case~II}
\end{figure}

In case II, there is a strong leading $W_L$-$W_R$ box diagram with
a $(t,t)$ pair of $t$ quark exchanges, with
$\xi_{d,t}^{LR}\xi_{d,t}^{RL}=A\lambda^2 (1-\rho+i\eta)
(-cs) e^{i(\sigma-\tau)}$, with phase $(\beta+\sigma-\tau)$,
from Eq.\ (\ref{Vi:Vii}b).
That is followed by a $(c,t)$ pair of order $m_t m_c \lambda$ whose
coefficients are thus about $1/6$ of the leading $(t,t)$ term.
We thus have $M_{d,12}^{LR} \sim M^{LL}_{d,12}$ at low $M_R$. 
$\chi^2$ contours for the ratio of the LR contribution
$|M_{d,12}^{LR}|$ from a $W_L$-$W_R$ 
 pair over the SM contribution
$|M_{d,12}^{LL}|$ as a function of $M_R$ are presented in 
Fig.\ \ref{relMd12:CaseII} for case II.
The upper 1-$\sigma$, 90\% CL, and 2-$\sigma$ contours show the largest 
LR contributions that still fit the six experimental constraints
with one degree of freedom for the three SM $V^L$ and two LR $V^R$
parameters, 
at $\chi^2 = 1.0$, $\chi^2 = 2.71$, and $\chi^2 = 4.0$, respectively. 
We see that the right-handed gauge boson $W_R$ can contribute
nearly as much to the \BBbard\ mixing
as the SM 
for $M_R$ out to at least $12$ TeV. 
In case II, the lower limit on $M_R$ is 600 GeV from a large
$120^\circ \le \delta \le 160^\circ$ region, and 900 GeV from
the normal $40^\circ \le \delta \le 80^\circ$ region.
The $V^R$ mixing angle $\theta$ prefers regions around $0^\circ$, 
$90^\circ$, and $180^\circ$ below 4 TeV, where the
leading $(t,t)$ contribution is smaller.  All values of the 
$V^R$ phase $\sigma$ are allowed above 2 TeV, while 
$\sigma \le 180^\circ$ is allowed below 2 TeV. 

\subsection{ $B_{s}$ systems}

\begin{figure}
\noindent\raisebox{30ex}%
  {$\displaystyle\frac{|M_{s,12}^{LR}|}{(|M_{s,12}^{LL}|+
|M_{s,12}^{LR}|)}$}
\includegraphics[scale=0.35]{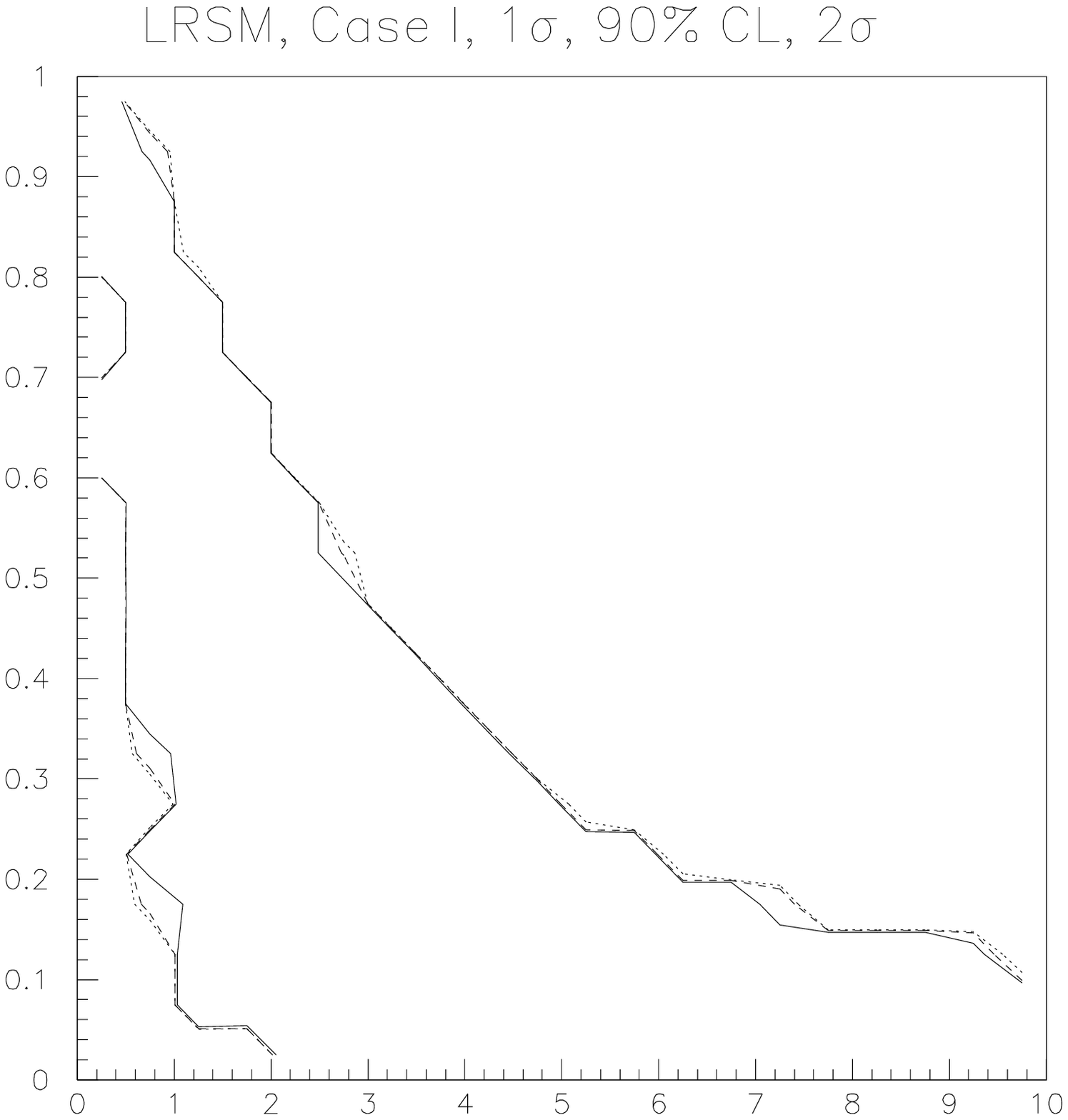}
\hspace*{10em}\mbox{$M_R$ (TeV)}
\caption{\label{relMs12:CaseI}%
  Plot of $|M_{s,12}^{LR}|/(|M_{s,12}^{LL}|+|M_{s,12}^{LR}|)$ as a function
  of $M_R$ for case~I}
\end{figure}

In \BBbars\ box diagrams for case I, in $M_{s,12}^{LR}$, four terms 
from $ (t,t)$,
$(c,t)$, $(t,c)$ and $(c,c)$ pairs
contribute. They are dominated by the
$(t,t)$ pair, with
$\xi_{s,t}^{LR}\xi_{s,t}^{RL}=A\lambda^2 cs e^{i(\sigma-\tau)}$.
The $\chi^2$ contours (again, for one degree of freedom) for 
the scaled ratio
$|M_{s,12}^{LR}|/(|M_{s,12}^{LL}|+|M_{s,12}^{LR}|)$ 
is presented
in Fig.\ \ref{relMs12:CaseI} for $M_R$ in the range of 0 to 10 TeV.  
This gives $M_{s,12}^{LR} \sim M_{s,12}^{LL}$ for
$M_R \le 2.5$ TeV, and $M^{LR}_{s,12} \sim 0.1 M^{LL}_{s,12}$ 
for $M_R = 10$ TeV.  The $W_{R}$ contribution to 
\BBbars\  mixing cannot be ignored  
if $M_R \le 5$  TeV. 
Although two $W_R$'s also appear in the \BBbars\        
box diagrams for case I,
with $(V_{ts}^{R*} V_{tb}^R)^2 = s^2 c^2 e^{2i(\sigma-\tau)}$, 
their effect is small 
due to the extra $\beta_R S(x_{tR})$ factor. 
In case I, the lower $M_R$ limit is about 100 GeV.  All $\sigma$
is allowed.  All $\theta$ are allowed down to 500 GeV, and below
that, the regions around $0^\circ$ and $180^\circ$ are preferred.

In case II, the $(t,t)$, $(t,c)$, $(c,t)$, and $(c,c)$ contributions
vanish from the structure of Eq.\ (\ref{Vi:Vii}b). 
The leading non-vanishing term in $M_{s,12}^{LR}$ comes from 
the $(t,u)$ pair, which is suppressed from the SM contribution
by $\lambda m_u/m_t$.
$W_{R}$ thus gives no effective contribution to
\BBbars\  mixing in case II.

\section{ {\boldmath $CP$} Violating Asymmetry  {\boldmath $\sin{2\beta}$} 
in {\boldmath $B^0$} Decays} 

The time dependent $CP$ violating phase  in $B \rightarrow \psi K_{S}$ decays 
is related to the mixing matrix element $M_{d,12}$ and the decay amplitudes as 
follows~\cite{nir},
\begin{equation} 
\sin{2\beta}  \equiv - {\rm Im} \left[ \frac{ M^{*}_{d,12}}{|M_{d,12}|}  
\frac{ A(\bar{B} \rightarrow \Psi K_{s})} {A(B \rightarrow \Psi K_{s})} 
\right].   
\end{equation}

In addition to the tree graphs,
the penguin diagrams, dominated by internal 
top-quarks, also 
contribute to  $B \rightarrow \psi K_{S}$ decays in case I.
The phase for the $W_{R}$ penguin amplitude,
$V_{tb}^{R}V^{R*}_{ts}=-cse^{i(\sigma-\tau)}$, is
exactly the same as
that  for the $W_{R}$ tree amplitude, 
$V_{cb}^{R}V_{cs}^{R*} = cse^{i(\sigma-\tau)}$.
 In the SM with $W_L$,
the penguin and tree phases are also equal. Accordingly,

\begin{eqnarray}
 A(\bar{B} \rightarrow \psi K_S) 
  & \propto & V_{cb}^{L}V_{cs}^{L*}(1-P) \nonumber \\
  & & + \beta_{gR} V_{cb}^{R}V_{cs}^{R*}(1-P'),
\end{eqnarray}
where $\beta_{gR} \equiv (g_R/g_L)^2 \beta_R$, and
$P$ and $P'$ are the ratios of the $W_{L}$ and $W_{R}$ penguin 
contributions over the tree amplitudes, respectively.
The first order approximation  $P \cong  P'
\propto \alpha_{s}{\rm ln}(m_{t}^{2}/m_{c}^{2})$ is applied to reach
the simplification of $P = P'$~\cite{kurimoto}. This makes
\begin{equation}
\sin{2\beta} = -{\rm Im}\left[ \frac{M^{*}_{d,12}} {|M_{d,12}|} 
\frac{ V_{cb}^{L}V_{cs}^{L*} + \beta_{gR} V_{cb}^{R}V_{cs}^{R*}}
{ V_{cb}^{L*}V_{cs}^{L} + \beta_{gR} V_{cb}^{R*}V_{cs}^{R} }\right].
\end{equation}

In case II, $W_R$ does not contribute to both the tree and penguin diagrams
in $B \rightarrow \psi K_{S}$ due to 
$V_{cb}^{R}V_{cs}^{R*}=0$ and $V_{tb}^{R}V_{ts}^{R*}=0$.
Therefore, we have
\begin{equation}                     
\sin{2\beta} = -{\rm Im}\left[ \frac{M^{*}_{d,12}}{|M_{d,12}|} 
\frac{ V_{cb}^{L}V_{cs}^{L*}}{V_{cb}^{L*}V_{cs}^{L} }\right].
\end{equation}

For both cases, since $\sin{2\beta}$ is a strongly constrained
input parameter, the results
are largely compatible with that 
obtained from the SM, whose range for the same data is
$0.63 \le \sin{2\beta} \le 0.82$, and no plot is shown. 
An earlier study of LRSM effects in $\sin{2\beta}$ is in
S. Nam \cite{nam}.

\section{ {\boldmath $\sin{\gamma}$} in 
{\boldmath $B$}$_s$ Decays}

Another asymmetry angle in $B$ meson systems is defined from
$B_s \to D_s^{+} K^{-}$ decays as~\cite{aleksan}
\begin{equation}
\sin{\gamma} \equiv
{\rm Im}\left(\frac{ M_{s,12}}{|M_{s,12}| } 
\frac{  A(B_{s} \rightarrow D^{+}_{s}K^{-})} 
  { A(\bar{B}_{s} \rightarrow D^{+}_{s}K^{-}) }\right).
\end{equation}
The penguin contribution is absent in both 
$B_{s} \rightarrow D^{+}_{s}K^{-}$ and  
$\bar{B}_{s} \rightarrow D^{+}_{s}K^{-}$ decays.
Because of the LRSM contribution, $\gamma$ as defined 
above is no longer an angle of the unitarity triangle.

The contributions from
$W_{R}$ to both decay modes in case I vanish since
$V_{ub}^{R*}V_{cs}^{R}=0$ and $V_{cb}^{R}V_{us}^{R*}=0.$ Therefore,
the $CP$ asymmetry for this decay mode can be simplified as 
\begin{equation}
\sin{\gamma} =
{\rm Im}\left(\frac{M_{s,12}} { |M_{s,12}|} 
\frac{V_{ub}^{L*}V_{cs}^{L}}{V_{cb}^{L}V_{us}^{L*}}
\bigg/ \left|\frac{V_{ub}^{L*}V_{cs}^{L}}{ V_{cb}^{L}V_{us}^{L*}}\right|
\right). 
\end{equation}

\begin{figure}
\noindent\raisebox{30ex}{$\sin\gamma$}
\includegraphics[scale=0.45]{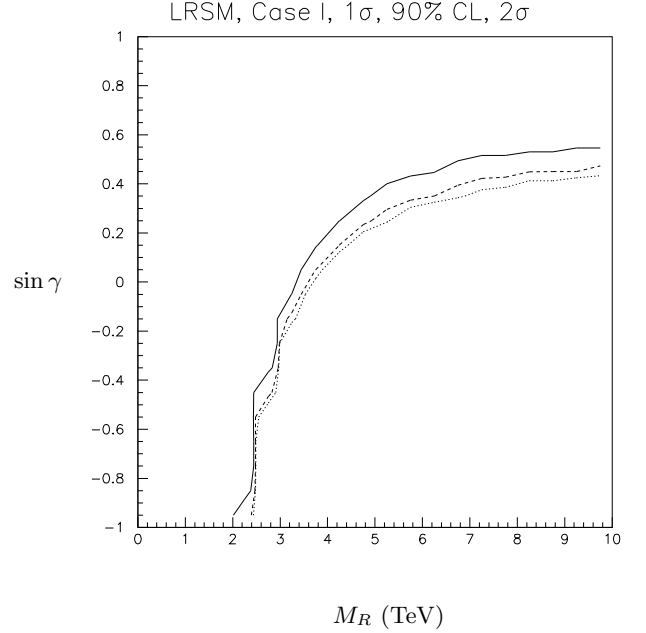}\newline
\hspace*{5em}\mbox{$M_R$ (TeV)}
\caption{\label{singamma:vs:Mr}%
  Plot of $\sin{\gamma}$ as a function of $M_R$ for case I}
\end{figure}

The allowed $\chi^2$ contours for the $CP$ asymmetry $\sin{\gamma}$ as 
a function of $M_R$ is shown in 
Fig.\ \ref{singamma:vs:Mr} for case I. 
We see that all values of  $\sin{\gamma}$ are allowed for
$M_R \le 2$ TeV, and that the 1-$\sigma$ range does not
limit itself to the SM range of 
$0.55 \le \sin{\gamma} \le 0.96$ until $M_R \sim 6$ TeV.

Because $W_{R}$ can contribute to $\bar{B}_{s} \rightarrow 
D_{s}^{+}K^{-}$ for case II, we have
\begin{eqnarray}
\sin{\gamma} = & {\rm Im} & \left( \frac{M_{s,12}} {|M_{s,12}|} 
            \frac{V_{ub}^{L*}V_{cs}^{L}} 
                  {V_{cb}^{L}V_{us}^{L*} +
                  \beta_{gR} V_{cb}^{R}V_{us}^{R*}}  \right.  \nonumber  \\
  &  &   \left. \bigg/
\left|\frac{V_{ub}^{L*}V_{cs}^{L}}{ V_{cb}^{L}V_{us}^{L*} 
+\beta_{gR} V_{cb}^{R}V_{us}^{R*}}\right|   \right).  
\end{eqnarray}
The $\sin{\gamma}$ values allowed in case II range from $0.64 \to 0.95$
at 1-$\sigma$ above $M_R = 7$ TeV, with a wider 1-$\sigma$ region
extending down to $\sin{\gamma} \ge 0.4$ for $M_R$ up to $7$ TeV
(graph not shown), which comes from a large $125^\circ \le \delta
\le 160^\circ$ region..  

\section{The {\boldmath $\sin{2\alpha}$} Asymmetry in
$B_d \to \pi \pi$}

Measurement on the asymmetry in $B_d \to \pi \pi$ can provide the other
$CP$ asymmetry, namely~\cite{nir},
\begin{equation} 
\sin{2\alpha}  \equiv - {\rm Im} \left[ \frac{ M^{*}_{d,12}}{|M_{d,12}|}  
\frac{ A(\bar{B} \rightarrow \pi \pi)} {A(B \rightarrow \pi \pi)} 
\right].   
\end{equation} 
                                 
In case I, there is no right handed tree nor penguin diagram for $B_d \to \pi \pi$
since $V_{ub}^{R}V_{ud}^{R*}=0$, $V_{cb}^R V_{cd}^{R*}=0$, 
and $V_{tb}^{R}V_{td}^{R*}=0$. 
On the other hand, the penguin pollution for this decay mode 
in the SM can be removed by isospin analysis~\cite{gronaulondon}. 
Consequently, we have

\begin{equation}
\sin{2\alpha} = {\rm Im}\left(\frac{M_{d,12}^*}{|M_{d,12}|}
\frac{V^{L*}_{ud} V^L_{ub}}
     {V^L_{ud} V^{L*}_{ub}}\right).
\end{equation}

\begin{figure}
\noindent\raisebox{30ex}{$\sin{2\alpha}$}
\includegraphics[scale=0.45]{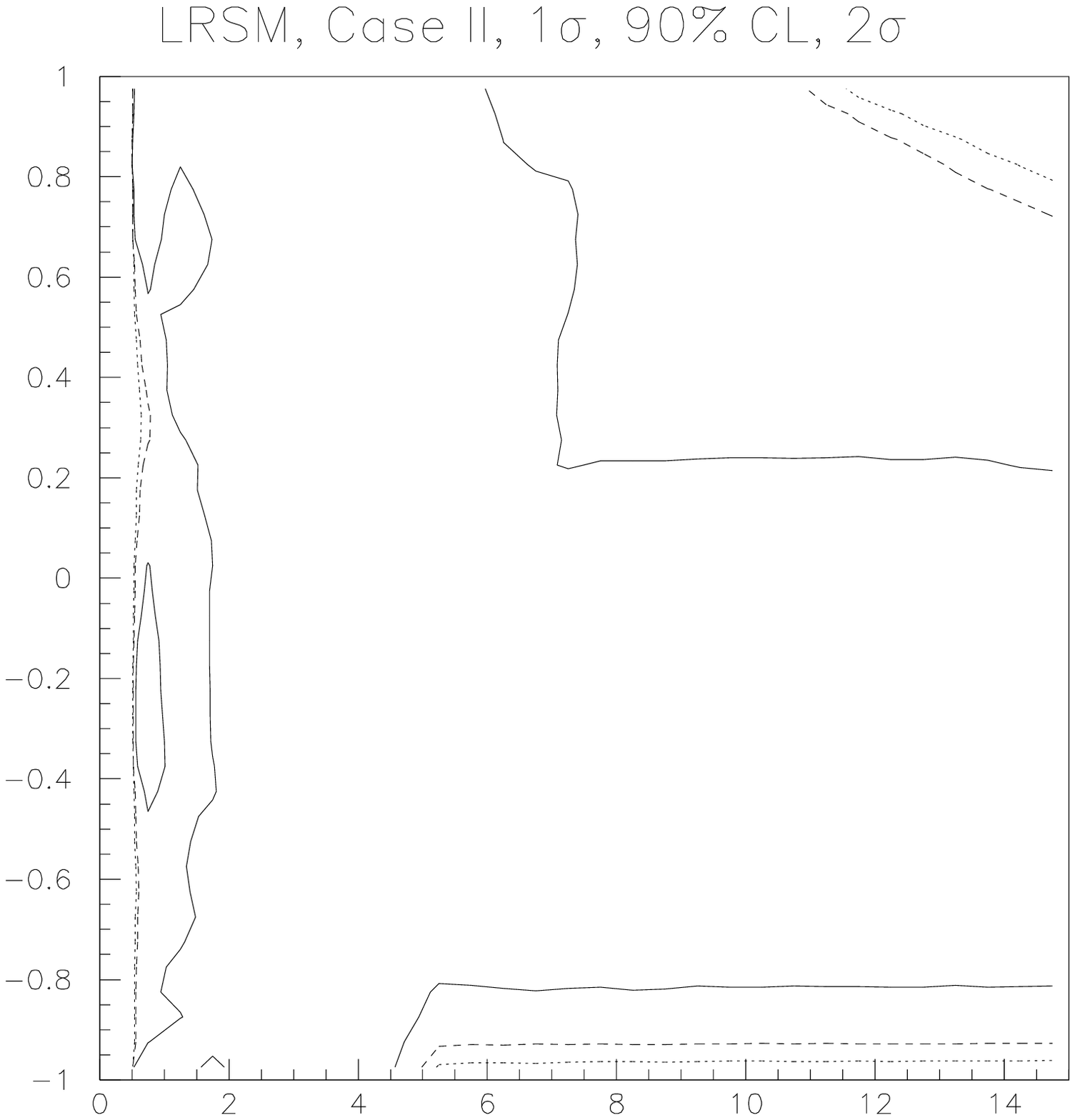}
\hspace*{5em}\mbox{$M_R$ (TeV)}
\caption{\label{sin2alpha:vs:Mr}%
  Plot of $\sin{2\alpha}$ as a function of $M_R$ for case~II}
\end{figure}

In case II, there are 
right handed $W_R$ penguin diagrams for $B_d \to \pi \pi$ from $b \to d$ through
virtual $t$ and $c$ quarks. 
The mixing product for these two processes, which are given in
$V_{tb}^{R}V^{R*}_{td}=-cse^{i(\sigma-\tau)}$ and 
$V_{cb}^{R}V_{cd}^{R*} = cse^{i(\sigma-\tau)}$, respectively,
are equal and opposite, but the amplitudes do not cancel since 
$m_t$ and $m_c$ are different.  They are included with the $W_L$ 
penguins to be isolated by isospin analysis.  We note that the 
right handed penguins might be appreciable since the left handed
penguins are suppressed by $A\lambda^3$. 
There is no right handed tree diagram in this case since $V_{ub}^R=0$.
We still use Eq.~(31) to calculate for case II, although again,
$\alpha$ is not the angle in the unitarity triangle in the LRSM.
Fig.\ \ref{sin2alpha:vs:Mr} shows the $\chi^2$ contours for 
$\sin{2\alpha}$ from $-1$ up to $1$ for $M_R \le 7$ TeV in case II,
and approximating the 1-$\sigma$
SM range for $\sin{2\alpha}$ of $-0.9 \le \sin{2\alpha} \le 0.33$ 
for larger $M_R$.  
There is a
large $125^\circ \le \delta \le 160^\circ$ region
that contributes out to $7$ TeV.
The positive $\sin{2\alpha}$ region
results from the larger $\delta$ region where 
$\alpha \le 90^\circ$ and $2\alpha \le 180^\circ$.

\section{The {\boldmath $\sin{2\phi}$}$_s$ Asymmetry in 
{\boldmath $B$}$_s$ Mixing}
\begin{figure}
\noindent\raisebox{30ex}{$\sin{2\phi_s}$}
\includegraphics[scale=0.40]{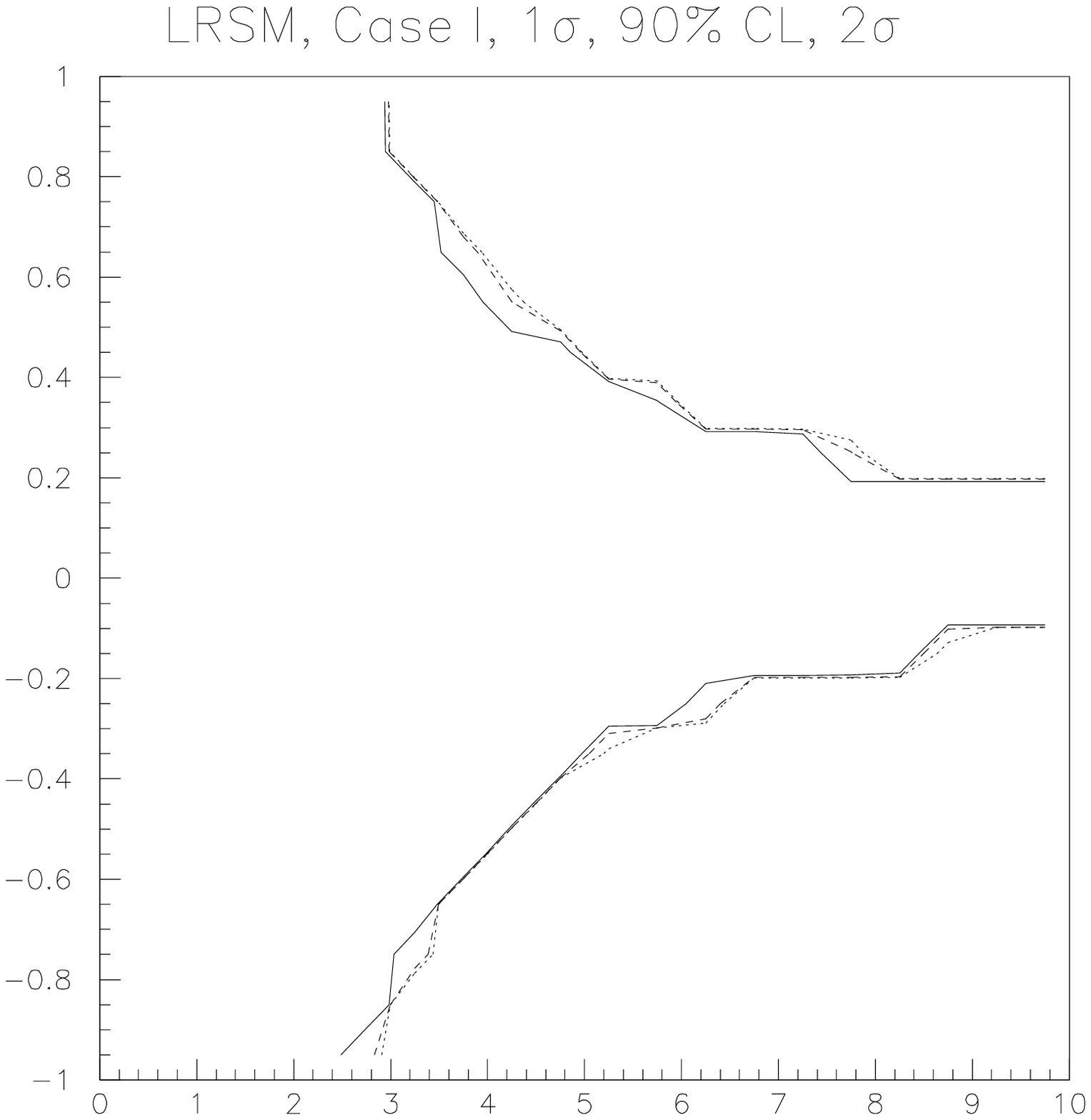}
\hspace*{5em}\mbox{$M_R$ (TeV)}
\caption{\label{sin2phis:vs:Mr}%
  Plot of $\sin{2\phi_s}$ as a function of $M_R$ for case~I}
\end{figure}

The mixing in $B_s$-$\bar B_s$ is the same as 
$\sin{2\phi_s}$, where $\phi_s$ is the small angle in
the SM ($b$-$s$) unitarity triangle.  Since this involves $B_s$ mixing,
it only appears in case I, and is given by~\cite{silvermanyao}
\begin{equation}
\sin{2\phi_s}= -{\rm Im}\left( \frac{M_{s,12}}{|M_{s,12}|}
\frac{V^{L*}_{cb}V^L_{cs}+\beta_{gR} V^{R*}_{cb}V^R_{cs}}
     {V^L_{cb}V^{L*}_{cs}+\beta_{gR} V^R_{cb}V^{R*}_{cs}}\right).
\end{equation}
Here, $W_R$ can also contribute to $\bar{b}\to\bar{c}c\bar{s}$
decays.  In the LRSM, with the asymmetry defined as above, 
$\phi_s$ is no longer the angle in the $V^L$ unitarity triangle.
The $\chi^2$ contours for $\sin{2\phi_s}$ in case I are shown in 
Fig.\ \ref{sin2phis:vs:Mr} and show 
$\sin{2\phi_s}$ from $-1$ to $1$ for $M_R$ up to $3$ TeV,
and above the SM range up to $8$ TeV.  
The SM 1-$\sigma$ $\sin{2\phi_s}$ range  
for the same data runs from 
$0.025 \le \sin{2\phi_s} \le .041$.  
This measurement could provide dramatic evidence for
new physics.
\section{$x_s$, the $B_s$ Oscillation Rate}
Since $x_s$ measures the $B_s$ oscillation rate, it 
is only affected in case I.  It is given by its ratio to
$x_d$ as~\cite{silvermanyao}
\begin{equation}
x_s = 1.034 x_d \left|\frac{M_{s,12}}{M_{d,12}}\right|,
\end{equation}
where $M_{s,12}$ contains the $W_R$ box diagrams, and $x_d = 0.77$.

\begin{figure}
\noindent\raisebox{30ex}{$x_s$}
\includegraphics[scale=0.45]{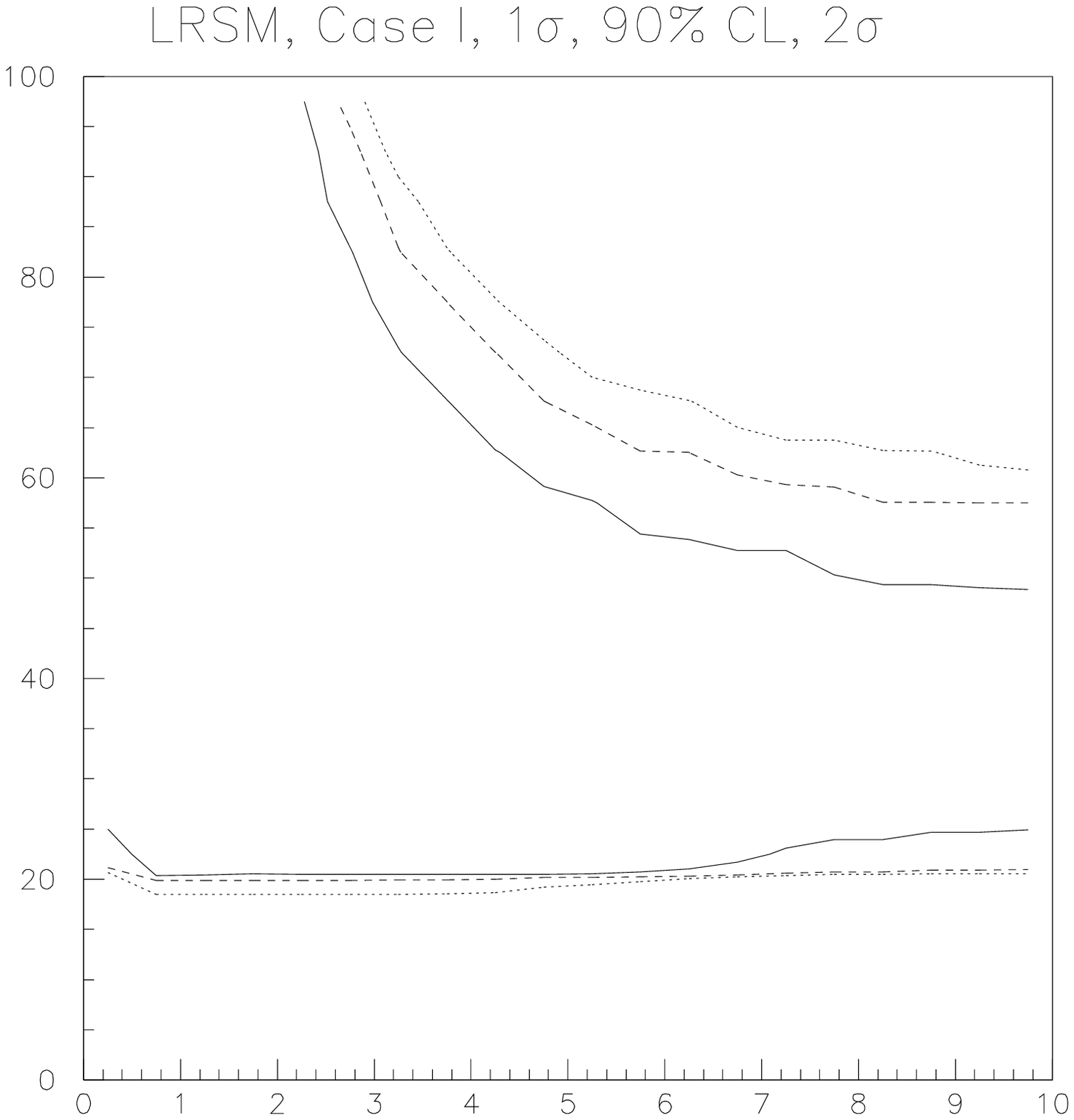}
\hspace*{5em}\mbox{$M_R$ (TeV)}
\caption{\label{xs:vs:Mr}%
  Plot of $x_s$ as a function of $M_R$ for case~I}
\end{figure}

The $\chi^2$ contours for $x_s$ are shown in
Fig.\ \ref{xs:vs:Mr} for case I.  
They show $x_s$ from 20 to greater than 100 for
$M_R \leq 2$ TeV.  There is no experimental upper limit to $x_s$.
In the LRSM, the order unity $V^R_{ts}$ matrix element for the $W_R$
exchange replaces the SM suppressed $V^L_{ts} = -A\lambda^2 = -0.04$
matrix element, giving low mass $W_R$ an initial advantage.
The SM 1-$\sigma$ range is  $24 \le x_s \le 53$, which is approached
for case I for $M_R$ above $5$ TeV.

\section{Conclusions}
In the Left-Right Symmetric Model, 
the right-handed quark mixing matrices can be parametrized into two 
cases as described in Eqs.~(\ref{Vi:Vii}), which provide
a reasonable lower limit for the $W_{R}$ mass \cite{langacker}.
We suppress the large contributions to  
$\epsilon_{K}$      
from the $W_{L}$-$W_{R}$ box diagram by effectively taking some 
parameters of
$V^{R}$ to vanish, as shown in Eqs.~(\ref{Vi:Vii:2}), so that
the quite small
experimental value of $\epsilon_{K}$ can be achieved and $W_{R}$
may give the most substantial effects on \CP\ asymmetries in $B$ 
decays~\cite{kurimoto} \cite{silvermanyao}.

In this paper we have
given the detailed calculations, as a function of $M_R$, 
of the total mixing matrix 
element $M_{q,12}$ and of its components
 $M_{q,12}^{LL}$, $M_{q,12}^{RR}$ and  $M^{LR}_{q,12}$, where
two $W_{L}$, two $W_{R}$ and a $W_{L}$-$W_{R}$ pair are exchanged in the
 box diagrams,
for both $B_d$ and $B_s$ systems.         
 The effects of $W_R$ 
are depicted by the ratios
 $|M_{q,12}^{LR}| /|M_{q,12}^{LL}|$ in both
$B_{d}$-$\bar B_{d}$  and $B_{s}$-$\bar B_{s}$ mixing, which
 are plotted in Figs.\ \ref{relMd12:CaseII} 
and \ref{relMs12:CaseI} for cases II and I, respectively.
In case I the LRSM shows significantly larger allowed regions
for $M_R \leq 6$ TeV 
in $B_s$ oscillation related $x_s$ and $\sin{\gamma}$, 
and beyond for $\sin{2\phi_s}$.
It effects $B_d$ asymmetries out to $1$ TeV.
In case II,  
the second quandrant $\delta$
region can affect the unitarity triangle angles and its vertex
for $M_R \leq 8$ TeV in the LRSM.  It
also affects $\sin{2\phi_s}$ and $x_s$ at 90\% CL beyond $8$ TeV.

Whereas $\sin{2\beta}$ as a function of $M_R$ is compatible with that 
obtained from the~SM, much larger allowed regions for $\sin{\gamma}$, 
$\sin{2\alpha}$, $x_s$, and $\sin{2\phi_s}$ are found when
the LR amplitudes are large, as stated above.

Consequently, measurements of the additional $CP$ violating asymmetries
beyond $\sin{2\beta}$ can provide interesting tests 
for the new physics given by the LRSM.


\end{document}